\documentclass[
 reprint,
 preprintnumbers,
 amsmath,
 amssymb,
 aps,
 prl,
 floatfix,
 superscriptaddress
]{revtex4-1}

\usepackage{color}
\usepackage{graphicx}
\usepackage{dcolumn}
\usepackage{bm}
\usepackage[version=3]{mhchem}
\usepackage{siunitx}
\usepackage{soul}
\def\DCP{\color[rgb]{0,0,0}} 
\begin{document}

\title{Kinetic stabilization of 1D surface states near twin boundaries in noncentrosymmetric BiPd}

\author{Chi Ming Yim}
\affiliation{SUPA and School of Physics and Astronomy, University of St Andrews, North Haugh, St Andrews, Fife, KY16 9SS, UK}%

\author{Christopher Trainer}%
\affiliation{SUPA and School of Physics and Astronomy, University of St Andrews, North Haugh, St Andrews, Fife, KY16 9SS, UK}%

\author{Ana Maldonado}%
\affiliation{SUPA and School of Physics and Astronomy, University of St Andrews, North Haugh, St Andrews, Fife, KY16 9SS, UK}%
\affiliation{Max-Planck-Institut f\"ur Festk\"orperforschung, Heisenbergstr.\ 1, 70569 Stuttgart, Germany}

\author{Bernd Braunecker}%
\affiliation{SUPA and School of Physics and Astronomy, University of St Andrews, North Haugh, St Andrews, Fife, KY16 9SS, UK}%

\author{Alexander Yaresko}
\affiliation{Max-Planck-Institut f\"ur Festk\"orperforschung, Heisenbergstr.\ 1, 70569 Stuttgart, Germany}

\author{Darren C.\ Peets}
\affiliation{Max-Planck-Institut f\"ur Festk\"orperforschung, Heisenbergstr.\ 1, 70569 Stuttgart, Germany}
\affiliation{Advanced Materials Laboratory, Fudan University, Shanghai 200438, China}

\author{Peter Wahl}%
\affiliation{SUPA and School of Physics and Astronomy, University of St Andrews, North Haugh, St Andrews, Fife, KY16 9SS, UK}%

\date{\today}

\begin{abstract}
{\DCP The search for one-dimensional (1D) topologically-protected electronic states has become an important research goal for condensed matter physics owing to their potential use in spintronic devices or as a building block for topologically non-trivial electronic states. Using low temperature scanning tunneling microscopy, we demonstrate the formation of 1D electronic states at twin boundaries at the surface of the noncentrosymmetric material BiPd.  These twin boundaries are topological defects which separate regions with antiparallel orientations of the crystallographic \textit{b} axis.  We demonstrate that the formation of the 1D electronic states can be rationalized by a change in effective mass of two-dimensional surface states across the twin boundary.  Our work therefore reveals a novel route towards designing 1D electronic states with strong spin-orbit coupling.}
\end{abstract}


\maketitle
{\DCP Electronic systems constrained to one spatial dimension lead to unusual physics and potential applications that cannot be realized in higher-dimensional electronic matter. The phenomena predicted and observed include the breakdown of the Fermi liquid and formation of a Luttinger liquid once electron interactions become important \cite{Haldane:1981}, as well as, in combination with superconductivity, Majorana bound states at the ends of a spin-filtered 1D electronic conductor with finite length, a possible path to topological quantum computation \cite{sarma_majorana_2015}.

The first potential observations of Majorana bound states have been achieved in heterostructures consisting of a 1D nanowire with strong spin-orbit interaction on a superconductor \cite{mourik_signatures_2012,das_zero-bias_2012,deng_anomalous_2012,rokhinson_fractional_2012}. 

Different routes can be used to create 1D electronic states: (1) through structural anisotropy in a bulk material, (2) through structural confinement of the electronic state, or (3) as edge states of topologically non-trivial electronic systems. Here we present a fourth route to create one dimensional electronic states, by demonstrating that such states can be generated through kinetic stabilization when two two-dimensional (2D) electronic states with slightly different effective masses come into contact. We explore this 1D state at the twin boundaries of noncentrosymmetric BiPd.}

Interest in noncentrosymmetric materials has been driven largely in the hope that they might host unconventional superconductivity. Theoretical studies of noncentrosymmetric superconductors have already suggested that twin boundaries in these materials exhibit unusual properties \cite{Arahata:2013gf,Aoyama:2014hg}. These predictions and other possible features of these 2D boundaries, and in particular their 1D intersection with a surface have not been explored experimentally so far.

The noncentrosymmetric superconductor BiPd \cite{Alekseevskii1952,Kheiker1953,zhuravlev_structure_1957} was recently rediscovered \cite{Joshi:2011bw} and has sparked substantial interest, not least due to availability of large single crystals that readily cleave. This makes them suitable for spectroscopic investigation by Scanning Tunneling Microscopy and Spectroscopy (STM/S) \cite{Sun:2015bj}, and by Angular Resolved Photoemission Spectroscopy (ARPES), which have revealed surface states with strong spin-orbit coupling \cite{Benia:2016df,thirupathaiah_unusual_2016} and intricate spin textures \cite{Neupane:2015uz}.

Previous reports have demonstrated that BiPd exhibits twin boundaries \cite{Sun:2015bj,Benia:2016df} even in high quality crystals with a residual resistivity ratio $R(300\mathrm K)/R(4\mathrm K)>100$. 

In this Letter, we focus entirely on the normal state and report on the discovery of 1D electronic states localized at twin boundaries in BiPd, studied by low temperature STM/S. The 1D electronic states are stabilized via a novel kinetic trapping mechanism whereby a change in effective mass of the two dimensional surface states between the domains leads to a locally attractive potential. In BiPd these 2D states are spin-split as a consequence of spin-orbit coupling, but the result is more general and applies also in systems without spin-orbit coupling. This mechanism provides a novel way to confine electronic states in low dimensional systems. 

Single crystals of BiPd were grown using a modified Bridgman-Stockbarger technique \cite{Peets:2014fn}.  STM/S measurements were performed using a home-built low-temperature STM operating at a base temperature of 1.8 K \cite{white_stiff_2011} as well as in an STM in a dilution refrigerator which can reach temperatures down to below 20mK\cite{singh_construction_2013}. Clean surfaces of BiPd single crystals were prepared by \textit{in situ} cleaving at 20 K and transfer of the sample into the STM. The STM tip was made from PtIr wire and conditioned by field emission on a gold single crystal. Bias voltages were applied to the sample with the tip at virtual ground. Differential conductance ($\mathrm{d}I/\mathrm{d}V$) spectra were recorded using a lock-in technique with a modulation frequency of 433 Hz. The surface electronic structure has been obtained using fully-relativistic linear-muffin-tin-orbital calculations (for details see refs.~\cite{Sun:2015bj, Benia:2016df}). 

\begin{figure}[ht]
\includegraphics[width=\columnwidth]{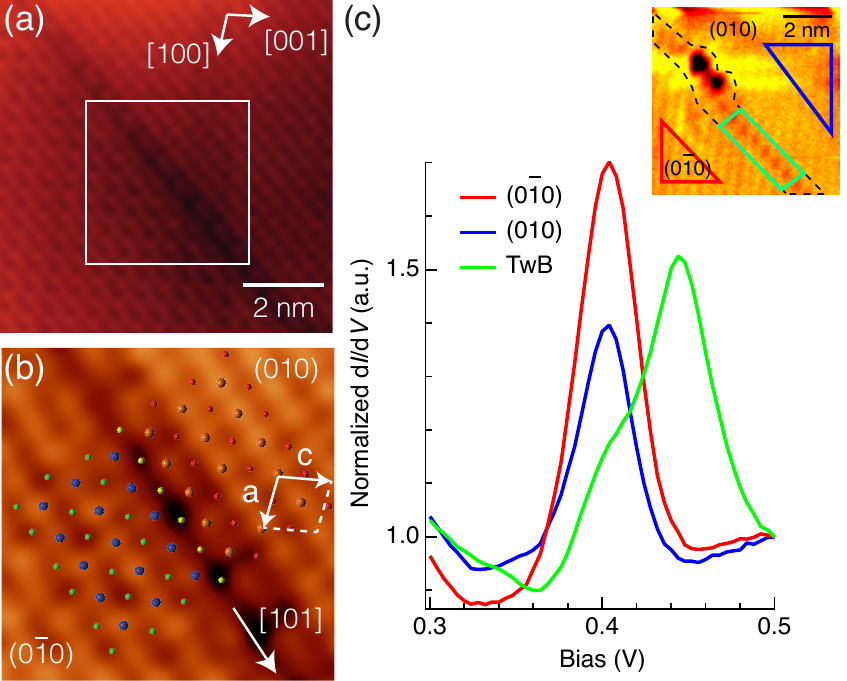}
\caption{(a) Topographic STM image of the surface of BiPd (($8.2\times8.2\, \mathrm{nm}^2$), $V_\mathrm t=80 \, \mathrm{mV}$, $I_\mathrm t=0.5\, \mathrm{nA}$).  (b) Zoomed-in image of the area marked with a white square in (a), overlaid with ball models for the $(0\overline{1}0)$ termination at the bottom left and the $(010)$ termination at the top right of the image respectively.  In the models, blue and orange spheres are Bi atoms, red and green spheres are Pd atoms, yellow 
spheres are Pd 
atoms which are connected to the neighboring Bi 
atoms from the opposite terminations on both side.  (c) Averaged $\mathrm{d}I/\mathrm{d}V$ spectra measured on the $(0\overline{1}0)$ and $(010)$ domains, as well as at the twin boundary ($V_\mathrm t=0.5\, \mathrm{V}$, $I_\mathrm t=50\, \mathrm{pA}$). Inset: STM image taken at a twin boundary ($7.9\times{}7.9$\,nm$^2$), marked by dashed lines.  Colored areas mark the regions over which the spectra in (c) were collected. STM images and spectroscopic data were obtained at 1.8~K.
}
\label{STM}
\end{figure}

Figure~\ref{STM}(a) shows a topographic STM image of a part of a freshly cleaved single crystal surface of BiPd, exposing a $(010)$/$(0\bar{1}0)$ termination. The scanned area shows two domains, one at the bottom left and the other at the top right, separated by a twin boundary.

The domains can be distinguished through the structural difference of the $(010)$ and $(0\overline{1}0)$ terminations \cite{Sun:2015bj}. By examining the arrangement of the surface atoms in both domains, and in particular the surface corrugation, we determine the domain at the bottom left to be the $(0\overline{1}0)$ termination, while that at the top right is assigned to a $(010)$ termination.

To analyze the atomic structure near the twin boundary we overlay two ball models, one on the $(0\overline{1}0)$ termination and one on the $(010)$ termination.  The resulting image, together with the proposed structure, is shown in Fig.~\ref{STM}(b) (for details of the model see ref.~\onlinecite{Supplement}).  As illustrated in Fig.~\ref{STM}(b), each of the Pd atoms in the surface layer at the twin boundary (yellow) is connected to four neighboring Bi atoms from the different domains on both sides. Similarly, the Bi atoms in the bottom layer are connected to the neighboring Pd atoms. The structural model of the twin boundary highlights that the coordination of the Bi and Pd atoms at the twin boundary remains the same as that within each of the domains, with only small variations in the local environment.

\begin{figure}[ht]
\includegraphics[width=\columnwidth]{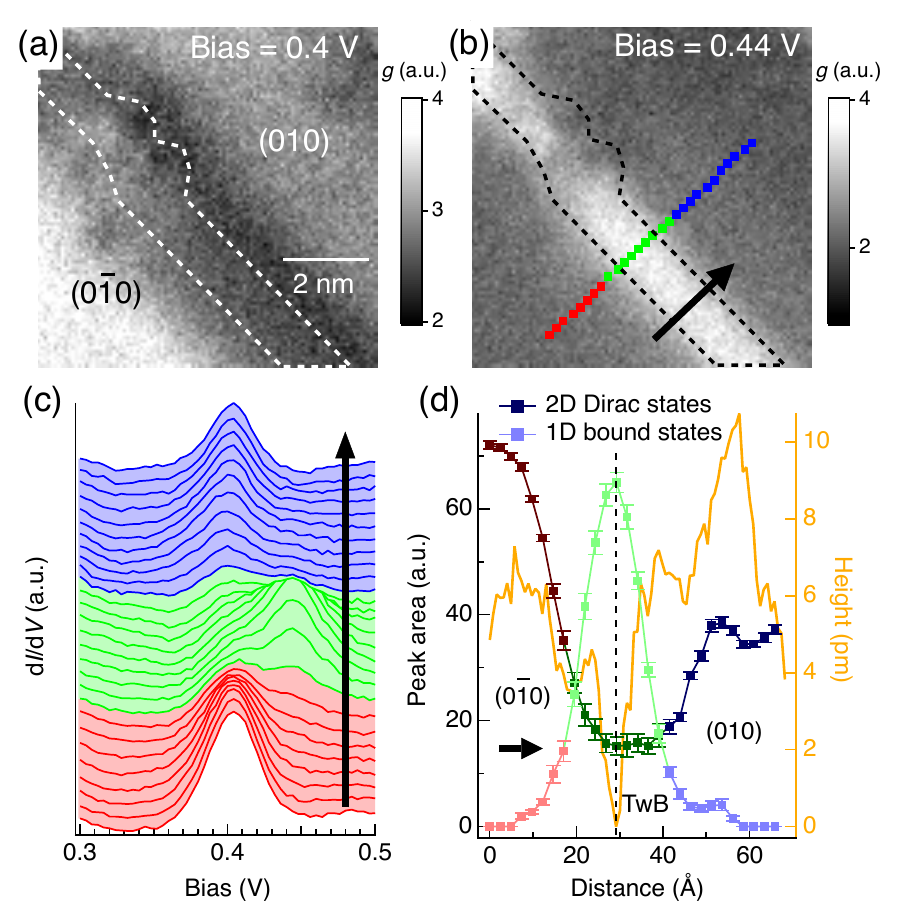}
\caption{(a-b) $\mathrm{d}I/\mathrm{d}V$ maps at bias voltages of (a) $0.4\, \mathrm V$ and (b) $0.44\, \mathrm V$, measured in the same field of view as the inset in Fig~\ref{STM}(c) ($7.9\times{}7.9\, \mathrm{nm}^2$).  Spectroscopy setpoint: $V_\mathrm t=0.5\, \mathrm{V}$, $I_\mathrm t=50\, \mathrm{pA}$.  Dashed lines enclose the twin boundary.  (c) $\mathrm{d}I/\mathrm{d}V$ spectra extracted from different positions at the left (red), middle (green), and right (blue) of the twin boundary, as indicated in (b).   Spectra are vertically offset for clarity.  (d)  Plot of peak areas of the 2D Dirac states (dark color) and 1D bound states (light color) versus position across the twin boundary.  Orange line is the line profile along the same positions across the twin boundary in the topographic image recorded simultaneously with the $\mathrm{d}I/\mathrm{d}V$ map.  Dashed vertical line marks the central position of the twin boundary.
}
\label{G-map}
\end{figure}

Previous ARPES and STM measurements have shown that the $(0\overline{1}0)$ and $(010)$ terminations are characterized by spin-split surface states which are located at different energies on opposite terminations \cite{Sun:2015bj,Benia:2016df}. The surface state observed at $\sim 0.4\, \mathrm{eV}$ is particularly prominent in tunneling spectra and has been shown to emerge in a directional bandgap opened by spin-orbit coupling, implying potential non-trivial topology \cite{Sun:2015bj}. In tunneling spectra this surface state is observed as a sharp peak at $\sim 0.4\, \mathrm{eV}$ above the Fermi energy [Fig.~\ref{STM}(c)], which is due to a van Hove singularity in the spin-split surface states. Calculations and ARPES \cite{Benia:2016df,yaresko_correct_2017} have shown that owing to the absence of inversion symmetry the dispersion of the Dirac states of the $(0\overline{1}0)$ and $(010)$ terminations are inequivalent. Furthermore, the surface states exhibit quite anisotropic effective masses in different directions in $k$-space.  

Figure~\ref{STM}(c) and Fig. S2 show tunneling spectra measured on the two sides of the twin boundary and, in addition, a spectrum measured right at a twin boundary. 
Spectroscopic data taken at a temperature of 30 mK (Fig. S2) show a clear signature of the superconducting gap, confirming that the twin boundary itself is superconducting. \cite{Supplement}. 
Tunneling spectra measured in a wider bias range reveal a strong bound state about $50\, \mathrm{meV}$ above the edge of the surface state at the twin boundary. Spatial maps of the differential conductance, and hence the local density of states, show the localized character of this bound state: Figures~\ref{G-map}(a) and \ref{G-map}(b) show $\mathrm{d}I/\mathrm{d}V$ maps acquired across the twin boundary shown in Fig.~\ref{STM}(b).  At bias voltage of $+0.4\, \mathrm{V}$ [Fig.~\ref{G-map}(a)], both of the $(0\overline{1}0)$ and $(010)$ terminations show a larger differential conductance than the twin boundary. However, at a bias voltage of $+0.44\, \mathrm{V}$ [Fig.~\ref{G-map}(b)], the twin boundary exhibits a substantially larger differential conductance than the other regions, signifying a larger density of states and the emergence of new electronic states at the twin boundary.

To gain further understanding regarding the nature of these twin boundary states, we examine $\mathrm{d}I/\mathrm{d}V$ spectra taken at different positions across the twin boundary.  As shown in Fig.~\ref{G-map}(c), the $\mathrm{d}I/\mathrm{d}V$ spectra extracted from the domains of opposite faces are characterized by a peak at bias voltage of $\sim{}0.4\, \mathrm{V}$, with that in the spectrum of the $(0\overline{1}0)$ termination appearing more pronounced and at slightly higher bias voltage compared to that in the spectrum of the (010) termination.  This is in very good agreement with previous reports \cite{Sun:2015bj}.  Approaching the twin boundary, the peak which we attribute to the surface state at $\sim{}0.4\, \mathrm{eV}$ becomes weaker in intensity, and a peak at $\sim{}0.44\,\mathrm{eV}$ emerges, reaching its maximum intensity at the center of the twin boundary. Even at the twin boundary center the surface state peak associated with the two terminations does not diminish completely [also see Fig. \ref{G-map}(d)].  

\begin{figure}[ht]
\includegraphics[width=\columnwidth]{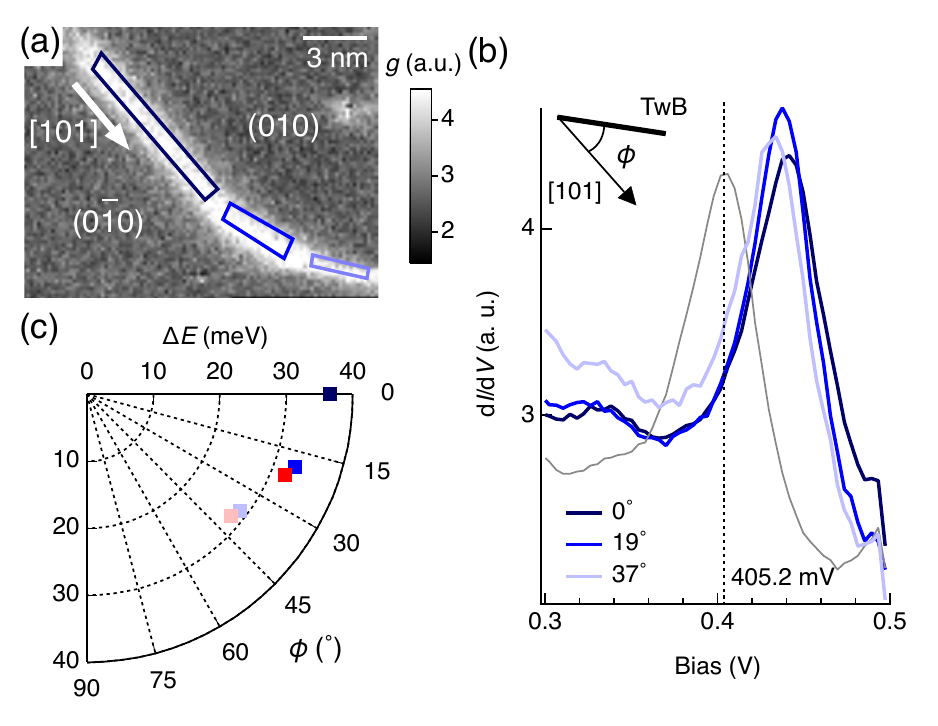}
\caption{(a) $\mathrm{d}I/\mathrm{d}V$ map recorded in the vicinity of a curved twin boundary separating domains of $(010)$ and $(0\overline{1}0)$ orientations, at bias voltage of 0.44\, V ($(16.6\times{}12.7)\,\mathrm{nm}^2$).  The twin boundary initially runs along the [101] direction from the top left, then deviates from this direction.  (b) Averaged $\mathrm{d}I/\mathrm{d}V$ spectra extracted from three different sections along the twin boundary marked with rectangles of different colors in (a).  The orientation of each section is represented by $\phi$, the angle of the twin boundary from the [101] direction (see inset).  An averaged spectrum extracted from the $(0\overline{1}0)$ domain (gray) is also included.  (c)  Polar plot of the energy position of the twin boundary bound state peak relative to that of the $(0\overline{1}0)$ domain surface state peak ($\Delta{}E$) as a function of the twin boundary orientation ($\phi$). Blue markers are the values determined from (b).  Red markers represent data points extracted from another dataset obtained with a different tip.  The corresponding spectra are shown in Fig. S3 of \cite{Supplement}. 
}
\label{Angular}
\end{figure}

The exact peak energy depends significantly on the orientation of the twin boundary: Figure~\ref{Angular}(a) shows a $\mathrm{d}I/\mathrm{d}V$ map taken at bias voltage of 0.44 V in the vicinity of a curved twin boundary, which initially runs along the $[101]$ direction and then bends away from it.  From the $\mathrm{d}I/\mathrm{d}V$ spectra extracted at three different positions along the twin boundary [Fig.~\ref{Angular}(b)], we find that the energy position of the bound state peak shifts to a lower bias voltage towards $E_{F}$ as the twin boundary orientation deviates from $[101]$.  A better illustration of this can be found in the angular plot in Fig.~\ref{Angular}(c).  The bound state peak is suppressed near defects (see Fig. \ref{G-map}) and completely absent in regions of the twin boundary with a high defect density (see Fig. S3), confirming the 1D nature of the twin boundary state.

\begin{figure}
\includegraphics[width=\columnwidth]{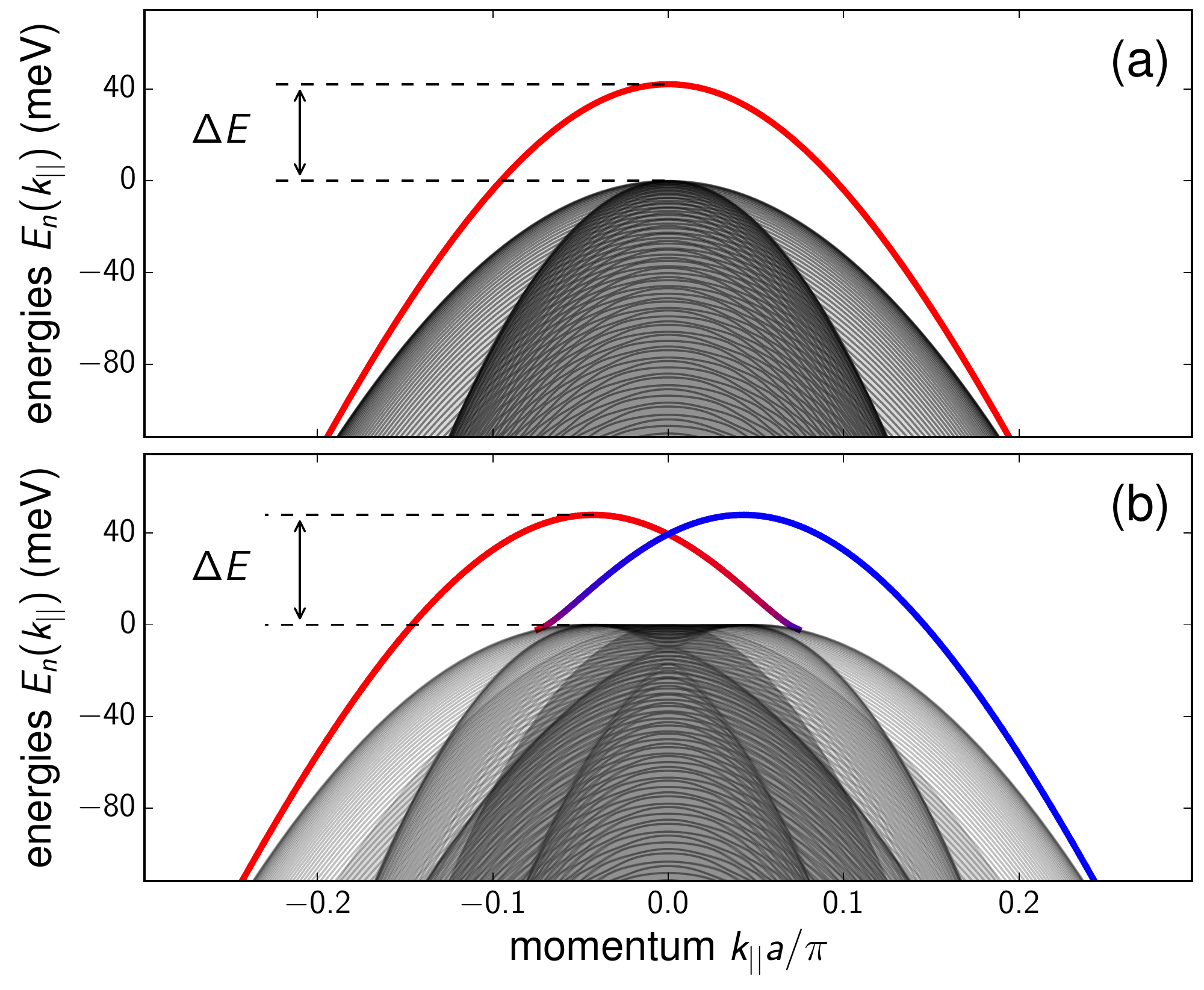}
\caption{%
Band structure of the tight binding model incorporating two different effective masses on both sides of the twin boundary, as a function of momentum $k_{||}$ parallel to the boundary, (a) without and (b) with spin-orbit interaction. When the top edges of the surface state bands on the two sides (shown in two different shades of gray) are approximately aligned a band of bound states (shown in red, red-blue) appears at the twin boundary, with maximal split-off energy $\Delta E$. In the bottom panel, spin-orbit interaction has been taken into account leading to a spin-splitting of the surface state bands (gray), but also of the bound state. The color coding in the bottom plot shows the in-plane spin polarization perpendicular to the twin boundary with red and blue for opposite spin directions.}
\label{fig:theory}
\end{figure}

We have explored different possible explanations for the origin of the twin boundary bound states. Since the bound states appear significantly away from the Fermi energy, superconductivity cannot cause the effect. The strong spin-orbit coupling in the material would suggest a mechanism based on the reversal of the symmetry breaking axis, however tight-binding calculations taking this into account do not support that this would lead to a bound state. The occurrence of a potential which localizes the surface state due to the structural distortion is possible, but seems unlikely: a potential dip that could create a bound state in this low dimensional electron gas \cite{simon_bound_1976} would very likely to be too small to create the large 50 meV offset of the bound state energy. The structural perturbation due to the twin boundary is minimal and does not affect the coordination of the atoms at the twin boundary. Also, such a mechanism would suggest that larger perturbations, such as step edges, would lead to a similar bound state, which is not something we observe --- rather the bound state immediately disappears when the two sides of the twin boundary separate. It turns out that the twin boundary states can be explained by a kinetic trapping mechanism based on the difference in effective mass on the two sides of the twin boundary. Table~\ref{effmass} shows that the effective masses, extracted from ab-initio calculations \cite{Sun:2015bj} are indeed significantly different for the 2D surface states at the different terminations. For the twin boundary shown in Fig.~\ref{STM}, the direction normal to it corresponds to the $\Gamma$-$S$ direction which shows a change in effective mass $\Delta m^\star=1.13$. The calculations further show that the surface states do not merge with 3D bulk states in the relevant energy range. It is therefore possible to describe the surface physics by a spatially varying effective mass across the twin boundary \cite{bendaniel_duke_1966}. This spatial change of kinetic energy takes the role of an effective potential, and if the latter becomes attractive over a region of space it can stabilize a band of bound states. To see that this is indeed sufficient we consider a minimal model including only the 2D surface bands in a tight binding description with the Hamiltonian $H = \sum_{\mathbf{r},\mathbf{a}, s,s'} t_{\mathbf{r},\mathbf{a}}^{s,s'} c_{\mathbf{r}+\mathbf{a},s}^\dagger c_{\mathbf{r},s'} - \sum_{\mathbf{r},s} \mu_\mathbf{r} c_{\mathbf{r},s}^\dagger c_{\mathbf{r},s}$. Here $\mathbf{r}$ runs over the positions of a 2D square lattice, $\mathbf{a}$ is the direction vector to the nearest neighbors, and $c_{\mathbf{r},s}$ and $c_{\mathbf{r},s}^\dagger$ are the electron creation and annihilation operators at site $\mathbf{r}$ with spin $s=\uparrow,\downarrow$. If $m_{A,B}$ are the effective masses in the regions $A=(0\overline{1}0)$ and $B=(010)$, we assume that the spin conserving hopping integral is $t_{\mathbf{r},\mathbf{a}}^{s,s} = t_{A,B} = \hbar^2/|m_{A,B}| a^2$ for hopping within each region (where $a$ is the effective lattice constant), and $t_{\mathbf{r},\mathbf{a}}^{s,s'} = -i \alpha_{A,B} (\mathbf{z} \times \mathbf{a}) \cdot \boldsymbol{\sigma}_{s,s'}$ is the spin-flip hopping by the spin-orbit interaction, with $\boldsymbol{\sigma}$ the vector of Pauli matrices, $\mathbf{z}$ the unit vector normal to the 2D plane, and $\alpha_{A,B} = \hbar^2 k^{so}_{A,B}/|m_{A,B}| a$. For simplicity we neglect the out of plane spin-orbit components. The values of $m_{A,B}$ and $k^{so}_{A,B}$ are listed in Table \ref{effmass}. We place the twin boundary parallel to one lattice direction, and for hopping across the boundary we use interpolated $t_I = \lambda t_A + (1-\lambda) t_B$ and $\alpha_I=\lambda \alpha_A+(1-\lambda)\alpha_B$, with $\lambda=0.8$ representing an asymmetry from the higher kinetic energy in region $A$. The chemical potentials $\mu_\mathbf{r}$ are constant in regions $A$ and $B$ and chosen such that the top edges of the bands are aligned at the same energy.
In Fig.~\ref{fig:theory} we show that this model leads indeed to a band of bound states. Figure~\ref{fig:theory}(a) shows that it is a purely kinetic effect and does not require spin-orbit interaction. In Fig.~\ref{fig:theory}(b) spin-orbit interaction is included and we observe that the bound states are spin-split in the same way as in purely 1D nanowires. We have assumed translational invariance along the twin boundary, for $500 \times 500$ sites with the boundary through the center, and used the effective masses from Table~\ref{effmass} ($\Gamma-S$ direction). A maximum split-off energy $\Delta E = 50$ meV in Fig.~\ref{fig:theory}(b) is obtained for the choice of $a=0.16$ nm. We note that $a$ corresponds to the length scale over which the effective mass changes through the boundary and not to the BiPd unit cell and indeed $\Delta E$ is sensitive to the choice of $a$ and $\lambda$. In particular at $\lambda < 0.5$ the bound states disappear, showing that the effective trapping potential at the boundary is just deep enough to bind a single and not multiple bands. This is confirmed by the experimental data but has to be included phenomenologically in the model through an asymmetric profile of the position-dependent effective mass.\\
\begin{table}[htb]
\begin{tabular}{@{}
lS[table-format=-2.2]S[table-format=-2.3]S[table-format=-2.2]S[table-format=-2.3]
@{}}
\hline
Direction & $m^*_{(010)}/m_\mathrm{e}$ & $k^\mathrm{so}_{(010)}$ $[\mathrm{\AA}^{-1}]$ & $m^*_{(0\overline{1}0)}/m_\mathrm{e}$ & $k^\mathrm{so}_{(0\overline{1}0)}$ $[\mathrm{\AA}^{-1}]$\\ 
\hline\hline
$\Gamma-S$        & -2.02 &  0.098 & -0.89 & 0.085\\
$\Gamma-S^\prime$ & -1.68 &  0.065 & -1.04 & 0.048\\
$\Gamma-X$        & -1.69 &  0.073 & -1.12 & 0.051\\
\hline
\end{tabular}
\caption{Relative effective mass $m^\star$ (in units of the bare electron mass) and spin-orbit momentum $k^{so}$ 
(in \AA$^{-1}$) of the surface state on opposite faces of BiPd in the $[010]$ direction, extracted from band structure calculations \cite{Sun:2015bj}. Because of the relatively large anisotropy of the band structure, $m^\star$ and $k^{so}$ are indicated for different high-symmetry directions.}
\label{effmass}
\end{table}
Our results identify a novel route to create 1D electronic states through the boundary between 2D states with slightly different electronic structure and a kinetic stabilization. In the case of BiPd, these surface states exhibit strong spin-orbit coupling, which is inherited by the twin boundary state. In order to explore this 1D state in transport, for spintronics applications or even to stabilize Majorana physics, it would need to be at the Fermi energy. While this is not the case here, the mechanism which leads to its formation does not depend on the Fermi energy.  Other non-centrosymmetric systems will exhibit similar 1D electronic states at twin boundaries, and it is likely that in some of these systems it can be tuned to the Fermi energy. The mechanism is applicable also in higher dimensions, implying that it can be used analogously to create 2D interface states.

\begin{acknowledgments}
CMY, CT and PW acknowledge funding from EPSRC through EP/I031014/1 and EP/L505079/1 and DCP acknowledges support from the National Natural Science Foundation of China (Project No.~11650110428).

Underpinning data can be obtained at Ref. \cite{Dataset} 
\end{acknowledgments}
\bibliography{BiPd-TwB.bib}
\pagebreak
\section{Supplemental Notes}
\setcounter{page}{1}
\setcounter{figure}{0}
\makeatletter
\renewcommand{\thefigure}{S\@arabic\c@figure}
\makeatother
\section{Further characterization of twin boundaries}
\begin{figure*}
\centering
\includegraphics[]{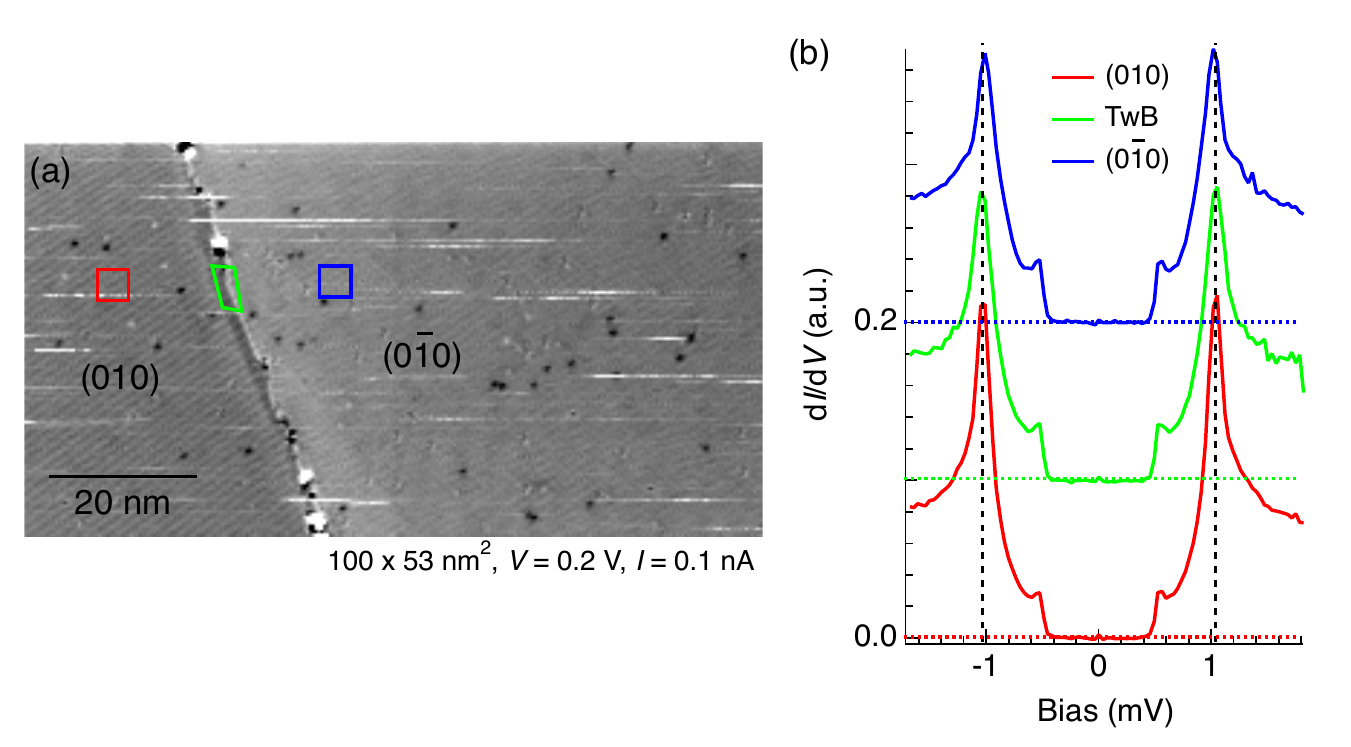}
\caption{(a) STM topographic image of a twin boundary separating two atomically flat domains of (010) and (0$\overline{1}$0) terminations of the crystal lattice.  (b) Averaged $\mathrm{d}I/\mathrm{d}V$ spectra recorded from different positions in the field-of-view in (a), which include the (010) (red) and ($0\overline{1}$0) (blue) terraces, and the twin boundary (green).   The spectra were taken at temperature of 30 mK using a superconducting tip, prepared by picking up a BiPd chunk from the surface, leading to two pairs of superconducting coherence peaks showing up in the spectra.  The spectra taken from different positions do not show any appreciable difference in the superconducting gap size.}
\label{SC}
\end{figure*}

\begin{figure*}
\centering
\includegraphics[]{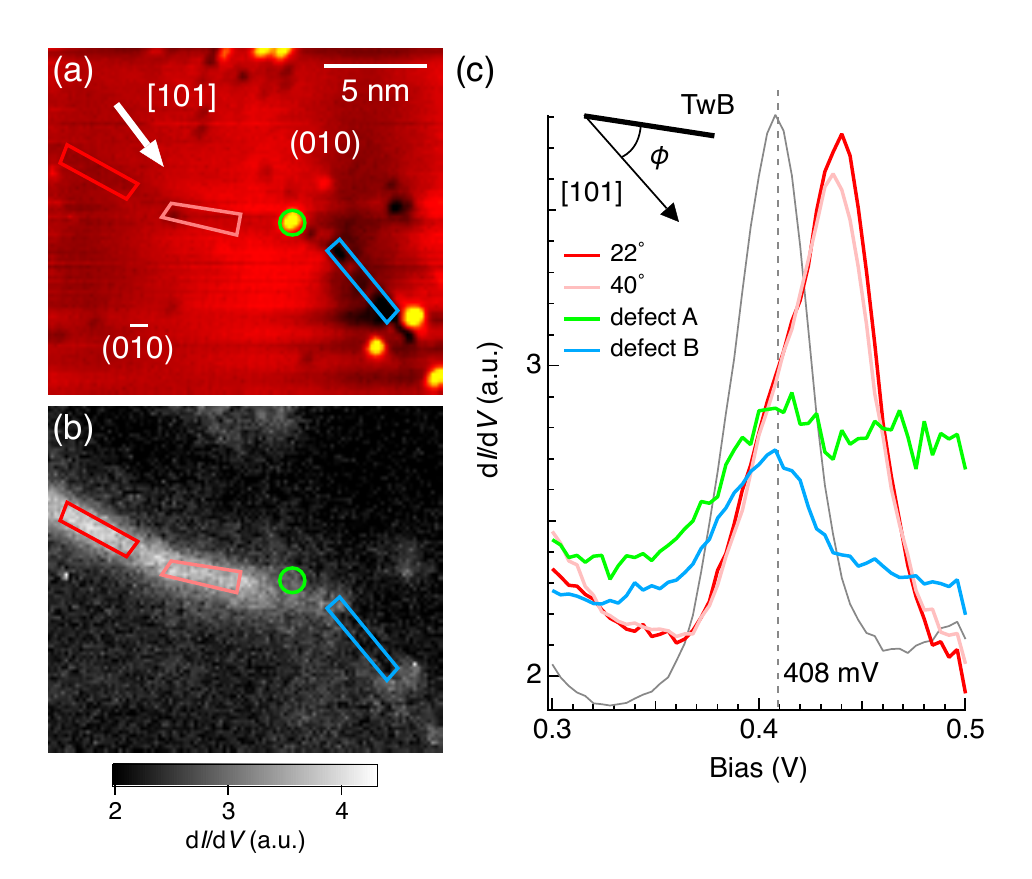}
\caption{(a) STM image taken at the lower part of the twin boundary shown in Fig.~3 in the main text ($19.4\times{}15.4$ nm$^2$).  A different tip was used.  (b) $\mathrm{d}I/\mathrm{d}V$ map slice taken simultaneously with the STM image in (a), at a bias voltage of 0.44 V.  (c)  Averaged $\mathrm{d}I/\mathrm{d}V$ spectra taken from the two different sections along the twin boundary as marked with red and pink rectangles in (a-b), respectively.  Each section is characterized by $\phi$, the angular separation between the section and the $[101]$ direction of the BiPd surface (see inset).  Spectra obtained from two different types of defects residing on the twin boundary [marked by green circles and blue rectangles in (a-b), respectively], as well as that from the $(0\overline{1}0)$ terrace, are also included.
}
\label{RedData}
\end{figure*}

Fig.~\ref{SC} shows a topographic image and spectroscopic data of the superconducting gap near a twin boundary.   The measurements were performed at a temperature of 30 mK.  In the field-of-view in Fig.~\ref{SC}(a), we recorded $\mathrm{d}I/\mathrm{d}V$ spectra at the twin boundary, as well as at the terraces as marked by different color boxes in Fig.~\ref{SC}(a).  Our spectroscopic data [Fig.~\ref{SC}(b)] show that the superconducting gap persists at the twin boundary with only minor changes. The data have been obtained with a superconducting tip.
\begin{figure*}
\centering
\includegraphics[]{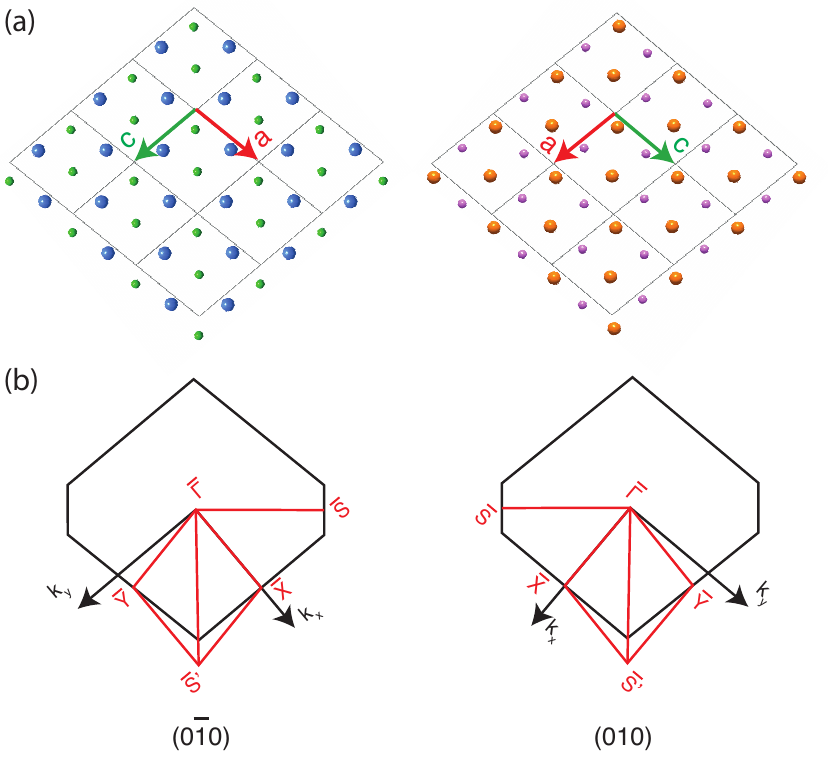}
\caption{(a) Schematic models of the (left) $(0\overline{1}0)$ and (right) $(010)$ terminations of the non-centrosymmetric BiPd.  Large blue and orange spheres are Bi atoms, small green and pink spheres are Pd atoms.  The models are aligned with their $[101]$ directions running parallel to each other, like the $(0\overline{1}0)$ and $(010)$ domains which are separated by a twin boundary as shown in Fig.\ 1 in the main text.  (b) Surface Brillouin Zones of the two terminations.}
\label{BrillouinZone}
\end{figure*}
Fig.~\ref{RedData} shows further spectroscopic data of the twin boundary bound state, and how it is affected by defects. These data were taken using a different tip.  We find that the twin boundary bound state is completely suppressed in the vicinity of defects as well as once the defect concentration becomes too high.

\pagebreak
\newpage
\end{document}